**Network organization of coopetitive genetic influences on cortical morphologies**

Subhadip Paul[1], Satyam Mukherjee[2], Sagnik Bhattacharyya[1,3] *

[1] Institute of Psychiatry, Psychology & Neuroscience, King's College London, UK.
[2] Indian Institute of Management, Udaipur, India.
[3] South London and Maudsley NHS Foundation Trust, Denmark Hill, Camberwell, London, UK

*Corresponding author
Department of Psychosis Studies, Institute of Psychiatry, Psychology & Neuroscience, King's College London, De Crespigny Park, London SE5 8AF, UK, Tel +44 207 848 0955, Fax +44 207 848 0976, Email: sagnik.2.bhattacharyya@kcl.ac.uk
`

*Main text*: 4824 words

Tables: 3

Figures: 2

*Supplementary Information:*

Supplementary Method

Supplementary Figure: 1

Supplementary References

**Abstract**

Brain can be represented as a network, where regions are the nodes and relations between the regions are edges. Within a network, co-existence of cooperative and competitive relationships between different nodes is called coopetition. Inter-regional genetic influences on morphological phenotypes (cortical thickness, surface area) of cortex display such coopetitive relationships. Here, we have represented these genetic influences as a network and shown that cooperative and competitive genetic influences on cortical morphological phenotypes follow distinct organization principles. Utilizing the theory of structural balance, we have shown that the pattern of collective regulation of cortical morphological phenotypes by cooperative and competitive genetic influences are overall bilaterally symmetric and such patterns of collective genetic regulation are similar to the principal modes of population variation of cortical morphological phenotypes. Finally, we have observed that the maximally and minimally imbalanced regions corresponding to the collective genetic regulation partially overlap with the cortical structural network hubs.

# Introduction

Human brain is a complex information processing system. Mathematical representations of the brain as a complex networked system has rendered original insights regarding the principles of organization of structural and functional brain networks (Bassett & Sporns, 2017; Bullmore & Sporns, 2009; Rubinov & Sporns, 2010; Stam & Reijneveld, 2007). These networks may be considered as the intermediate phenotypes mediating the effect of genetic and related molecular systems on behaviour (Bassett & Sporns, 2017). For example, genetic variations may be associated with altered structural and/ or functional brain networks, which in turn may alter behaviour and vice versa. Within a networked system, a node may interact cooperatively with some nodes and competitively with other nodes (Hu & Zheng, 2014). Coexistence of cooperative and competitive relationships between the nodes within a network is defined as coopetition (Brandenburger & Nalebuff, 1996; Hu & Zheng, 2014). Such coopetitive relationships are often observed between various members within networks of human relationships (Wasserman & Faust, 1994), social networks of animals (Ilany, Barocas, Koren, Kam, & Geffen, 2013), and, networks of business organizations (Bengtsson & Kock, 2000).

The complex process of cortical patterning into functionally and cyto-architecturally distinct regions in the developing mammalian cerebral cortex, known as arealization, is orchestrated by intrinsic genetic mechanisms and thalamocortical pathways relaying extrinsic information (O'Leary, Chou, & Sahara, 2007). Broadly similar patterns of genetic influences observed in adult human and non-human mammalian brains suggest conservation of cortical patterning mechanism (Chen et al., 2011). Neuroimaging based twin study designs offer the opportunity to non-invasively examine aggregate genetic influence on the relative expansion or contraction of two anatomical regions of the human brain (Chen et al., 2012; Chen et al., 2011). The extent to which two anatomical regions share common genetic influence (degree of pleiotropy) can be estimated using genetic correlation (Almasy, Dyer, & Blangero, 1997) between the morphological phenotypes (cortical thickness, surface area) of those regions (Chen et al., 2012; Chen et al., 2011). Positive genetic correlation between two anatomical regions would imply that genetic influences which result in the expansion of one region would also result in the expansion of the other region (Chen et al., 2011). On the other hand, negative genetic correlation would suggest that the genetic influences which expand one region, would have an opposite effect on the other region and result in its contraction (Chen et al., 2011). Animal studies showing expansions of anterior and lateral regions of the cortex in *Emx2* null mice supports the idea of positive genetic influences between brain regions (Mallamaci, Muzio, Chan, Parnavelas, & Boncinelli, 2000). Studies involving mutant mice have also shown that regulatory genes, such as the *Pax6* and *Emx2* that control neocortical arealization have opposing effects in caudal and rostral brain regions (Bishop, Goudreau, & O'Leary, 2000), supporting the notion of negative genetic influences between brain regions. We have denoted those positive and negative genetic influences as inter-regional cooperative and competitive genetic influences on morphological phenotypes respectively. Morphological phenotype of a cortical region may have positive genetic correlations with some regions and negative correlations with other regions (Chen et al., 2012; Chen et al., 2011). Coexistence of such cooperative and competitive inter-regional genetic influences on morphological phenotypes suggest that inter-regional genetic influences in the human cerebral cortex are coopetitive.

We have represented inter-regional coopetitive genetic influences on cortical morphological phenotypes as signed networks where cortical regions are the nodes and positive and negative genetic correlations denote the weighted links between the nodes. However, whether co-operative and competitive genetic influences on cortical morphological phenotypes follow the same network organization principles has remained elusive. Therefore, we have compared network properties of co-operative and competitive inter-regional genetic influences on morphological phenotypes to derive new insights into the principles of morphological organization of human cerebral cortex at the level of genetic regulation.

In the context of triadic social relationships between individuals, the theory of structural balance (F. Heider, 1946; Fritz Heider, 1958) has considered competitive relationships between the individuals as a potential source of conflict or imbalance in a system (detailed in Materials and methods). Extension of the theory of structural balance (F. Heider, 1946; Fritz Heider, 1958) to the context of networked systems (Cartwright & Harary, 1956; Estrada & Benzi, 2014) has provided an opportunity to examine the global and local network organization collectively shaped up by the co-operative and competitive relations between the nodes. The states of such network organization can be characterized in terms of the extent to which they are balanced or imbalanced (Estrada & Benzi, 2014). We have utilized the notion of structural balance to examine the states of collective regulation of global and local morphological organization of cerebral cortex by the inter-regional cooperative and competitive genetic influences. Subsequently, we investigated whether there is correspondence between such spatial patterns of collective genetic regulation and the principal modes of variation of these morphological phenotypes within the participants. Finally, we examined the similarities and dissimilarities between the coopetitive genetic network organization underlying cortical thickness and surface area which are controlled by distinct genetic (Panizzon et al., 2009; Winkler et al., 2010) and cellular processes (Chenn & Walsh, 2002). To our knowledge, these have not been explored before.

In this article, we have used structural magnetic resonance imaging (sMRI) data from genetically confirmed twins and siblings to investigate the organization of genetic networks underlying morphological phenotypes of cerebral cortex in human.

**Results**

Mean degrees of co-operative and competitive genetic influences on regional cortical thickness were not significantly different (p = 0.61) (Table 2). Mean degrees of the genetic influences on regional cortical surface areas were also not significantly different (p = 0.36) (Table 2). Results from comparison of the network properties of co-operative and competitive genetic influences on both cortical thickness and surface area are reported below.

**Assortative co-operations and disassortative competitions**

Global assortativity of the co-operative genetic influences on both cortical thickness (p = 5 × $10^{-5}$) and surface areas (p = 5 × $10^{-5}$) were significantly higher than the random network (Table 1). Positive values of global assortativity of co-operative genetic influences on both cortical thickness and surface area suggest that the cortical regions with high number of co-operative genetic influences tend to associate with other cortical regions with high number of co-operative influences. Negative values of global assortativity measures of the competitive genetic influences on both cortical morphological phenotypes suggest that these competitive influences were dis-assortative (Table 1). Competitive genetic influences on regional cortical thickness were significantly (p = 5 × $10^{-5}$) less disassortative while those influences on surface areas were significantly (p = 5 × $10^{-5}$) more disassortative than corresponding random networks (Table 1). Disassortative mixing pattern of competitive genetic influences suggest that the cortical regions with a high number of competitive genetic influences tend to associate with a region with low number of competitive influences.

**More cohesive co-operative genetic influences than competitive genetic influences**

Co-operative genetic influences on both regional cortical thickness (p = 5 × $10^{-5}$) and surface area (p = 5 × $10^{-5}$) were significantly more clustered than the corresponding random networks (Table 1). On the other hand, the competitive genetic influences on these cortical morphological phenotypes were significantly (p = 5 × $10^{-5}$) less clustered than the random networks (Table 1). Mean nodal clustering coefficients of co-operative genetic influences were significantly higher than the competitive genetic influences on both cortical thickness (p = $10^{-4}$) (Table 2) and surface area (p = $10^{-4}$) (Table 2). Our results suggest that the co-operative genetic influences on both morphological phenotypes are more cohesive than the competitive genetic influences.

**More diverse competitive genetic influences than co-operative genetic influences**

Genetic networks underlying both cortical thickness (p = 5 × $10^{-5}$) and surface area (p = 5 × $10^{-5}$) were significantly more modular than the random networks (Table 1). Such modular network architecture contains the effects of local genetic or environmental perturbation mostly within the module where it originates and thus enhances the robustness against the widespread effects of the perturbation (Sporns & Betzel, 2016). Significantly higher values of mean nodal participation coefficients of competitive genetic influences than the co-operative genetic influences underlying both cortical thickness (p = 0.0013) and surface area (p = $10^{-4}$) (Table 2) suggest that on average, competitive genetic influences were more diverse than the co-operative genetic influences on both cortical morphological phenotypes. The values of modularity and participation coefficients depend upon the choice of resolution parameter ($\gamma$) during modularity estimation. These results were obtained with $\gamma$ = 0.5. However, conclusions remained unchanged (p < 0.004) for other considered values of $\gamma$ (See Supplementary Figure 1).

**Organization of structural balance of cortical genetic networks**

Genetic networks underlying thickness (p = 5 × $10^{-5}$) and surface area (p = 5 × $10^{-5}$) had significantly lower values of global imbalance measures than the corresponding random networks (Table 1). These observations suggest that the states of collective regulation of cortical morphological phenotypes by cooperative and competitive genetic influences were more globally balanced than comparable random networks.

In the context of genetic network underlying cortical thickness, means of nodal imbalance at left (mean ± std = 0.395 ± 0.209) and right (mean ± std = 0.399 ± 0.196) hemispheres were not significantly different (p = 0.94). Similarly, for genetic network underlying surface area, means of nodal imbalance at left (mean ± std = 0.257 ± 0.085) and right (mean ± std = 0.253 ± 0.084) hemispheres were also not significantly different (p = 0.87). These observations suggest that the states of collective regulation of thickness and surface area of cerebral cortex by coopetitive genetic influences were on average bilaterally symmetric.

State of collective regulation of cortical thickness by coopetitive genetic influences was maximally imbalanced at bilateral superior, middle and inferior temporal and precentral gyri and minimally imbalanced at bilateral inferior frontal (parstriangularis, parsorbitalis) and lingual gyri (Figure 2a). State of collective regulation of surface area by coopetitive genetic influences was maximally imbalanced at the left transverse temporal gyrus/Heschl's gyrus and the right supramarginal gyrus and minimally imbalanced at the left caudal anterior cingulate and right posterior cingulate cortex (Figure 2b). Middle and superior temporal gyri and anterior and posterior cingulate cortices are some of the structural network hub regions of the cerebral cortex (van den Heuvel & Sporns, 2013). Maximally and minimally imbalanced regions corresponding to the collective genetic regulation therefore seems to partially overlap with the structural network hubs of the cortex.

**Genetic structural balance and modes of phenotypic variations**

In a principal component (PC), cortical regions with higher absolute values of loading play a greater role in explaining the phenotypic variation on that PC (Figures 2c, 2d). Positive and negative values of the regional loadings imply their contributions to the PC in opposite directions in explaining phenotypic variation. In the context of cortical thickness, we observed significant positive correlation ($\rho = 0.69$, $p < 10^{-16}$) between the nodal measures of genetic structural imbalance and the loadings of principal component 5 (PC5) of cortical thickness variation within the participants. PC5 explained 3.27% of cortical thickness variation within the participants. Each of the first four PCs (variance explained: 11.73%, 10.15%, 5.03%, 4.15%) did not correlate significantly with the nodal genetic imbalance measures ($p > 0.05$). In the context of surface area, we observed significant negative correlation ($\rho = -0.42$, $p = 0.0004$) between the nodal measures of genetic structural imbalance and the loadings of PC5 of surface area variation. PC5 explained 3.36% surface area variation within the participants. Each of the first four PCs (variance explained: 8.88%, 5.82%, 4.41%, 3.81%) did not correlate significantly with the nodal genetic imbalance measures ($p > 0.05$) for cortical surface area. Therefore, our results suggest a significant similarity between the patterns of collective genetic regulation of cortical morphological phenotypes and some of the principal modes of variation of those phenotypes within the adult participants. However, as cortical morphological phenotypes in the adult human brain are influenced by both genetic and environmental factors(Panizzon et al., 2009), environmental influences are also expected to play an important role in phenotypic variation within participants.

**Dissimilar genetic network organizations underlying thickness and surface area**

Significant (FDR, $q < 0.05$) heritability of thickness ($h^2$ range: 0.14 to 0.68) and surface area ($h^2$ range: 0.12 to 0.79) of all cortical regions confirmed previous evidence (Panizzon et al., 2009; Winkler et al., 2010) of genetic control of thickness and surface area of all cortical regions. Further, the lack of significant genetic correlations between the thickness and corresponding surface area of all cortical regions also confirmed the previous evidence (Panizzon et al., 2009; Winkler et al., 2010) that different genetic factors control cortical thickness and surface area (FDR, q > 0.05).

Correlations of nodal degree, assortativity, clustering coefficient and participation coefficient of cooperative genetic influences between cortical thickness and surface area were not significant (p > 0.05, Table 3). Correlations of these nodal properties of competitive genetic influences between cortical thickness and surface area were also not significant (p > 0.05, Table 3). Furthermore, nodal imbalance properties of genetic networks underlying cortical thickness and surface area were not significantly correlated (p > 0.05, Table 3). These observations suggest that cortical regions play very different roles in the contexts of number of links, assortativity, clustering, diversity and nodal structural imbalance of genetic influence networks underlying cortical thickness and surface areas. Distinctive genetic influences on cortical thickness and surface area may underlie these dissimilarities in network organizations.

## Discussions

Our results furnish new insights on the differences in the organizational principles of cooperative and competitive genetic influences on thickness and surface area of the human cerebral cortex. In summary, co-operative genetic influences are more cohesive/clustered and diverse than competitive genetic influences. Co-operative genetic influences are assortative while competitive influences are disassortative. Assortative mixing pattern of co-operative genetic influences renders the corresponding highest degree nodes resilient to perturbation (Newman, 2002). On the other hand, disassortative mixing pattern of competitive genetic influences underlying the cortical morphologies suggests that the corresponding highest degree nodes are vulnerable to perturbation (Newman, 2002). Modular organization of coopetitive genetic networks largely restricts the effect of local genetic and environmental perturbations within the module where they initiate and improve the network robustness against those perturbations (Sporns & Betzel, 2016). Dissimilarities in the network organization between cortical thickness and surface area could be related to the influences of different sets of genetic factors on these morphological phenotypes.

We have found that the spatial patterns of collective genetic regulation of both morphological phenotypes by coopetitive genetic influences are on average bilaterally symmetric. However, the patterns of collective genetic regulations on the left and right hemispheres are not mirror reflections of each other. Such differences at the local level may relate to lateralization of brain functions. Significant correspondence between such spatial collective genetic regulation pattern with the loadings of the principal mode of morphological phenotypic variation implies that such collective genetic regulation may partly contribute to phenotypic variation within participants. However, as cortical thickness and surface area are influenced by both genetic as well as environmental factors, it is likely that environmental factors would also have an important effect.

There are some striking similarities between the global organizational principles of coopetitive genetic influence networks of the human cerebral cortex and human relationship networks in online social networks. As we observed with the genetic networks, co-operative human relationships in online social networks are also more cohesive than competitive relationships (Szell, Lambiotte, & Thurner, 2010). These co-operative human relationships are assortative while competitive relationships are dis-assortative (Ciotti, Bianconi, Capocci, Colaiori, & Panzarasa, 2015). Furthermore, similar to the genetic networks, online social networks are also structurally more global balanced than the corresponding random networks (Estrada & Benzi, 2014).

Maximally and minimally imbalanced regions of coopetitive genetic networks underlying cortical morphological phenotypes also overlap partially with the brain structural network hubs, which are implicated in different brain disorders (Crossley et al., 2014). Previous studies have reported shared genetic influences between cortical morphological phenotypes and schizophrenia (Bakken et al., 2011), bipolar disorder (Bootsman et al., 2015), multiple sclerosis (Matsushita et al., 2015), cannabis addiction (Paul & Bhattacharyya, 2018) and development (van Soelen et al., 2012). Investigation of the alterations of coopetitive genetic network organization underlying cortical morphological phenotypes in those conditions may furnish new disorder-specific insights that may be associated with the disease state or lead to vulnerability to those conditions. In summary, we view our results as a benchmark, which could be used in the future studies to examine the disruptions in coopetitive genetic network organization in brain diseases and under experimentally-induced perturbations.

## Materials and Methods

**Participants and anatomical images**

We analysed data from 593 siblings (320 females) from the S1200 release of Human Connectome Project (HCP) database (https://db.humanconnectome.org/) in this study. Participants had a mean (standard deviation) age of 28.84 (3.51) years ranging between 22 to 36 years. Mini Mental State Examination (MMSE) score ranged between 24 to 30. There were 138 monozygotic, 79 dizygotic and 108 non-twin gender-matched sibling pairs. Genotyping was carried out employing customized microarray chip using DNA extracted from blood/saliva of participants to confirm biological family relationships (shared parents and zygosity of twin pairs). Laterality quotient of the participants was assessed using Edinburgh Handedness Inventory, and could vary from +100 (completely right-handed) to -100 (completely left-handed)(Schachter, Ransil, & Geschwind, 1987). In vivo T1 weighted Magnetic Resonance (MR) images of the brains of the participants were acquired using a customized Siemens 3T Skyra scanner fitted with 32 channel head coil. 3D MPRAGE sequence with the following pulse sequence parameters were used: matrix size =320, field of

view = 224 mm, number of sagittal slices = 256, voxel dimension = 0.7 mm isotropic, TR/TE/T1/echo spacing = 2400 ms/ 2.14 ms/ 1000 ms/ 7.6 ms, flip angle = 8 degree, GRAPPA factor = 2 and bandwidth = 210 Hz/ pixel (Van Essen et al., 2012). All experiments were conducted following relevant guidelines and regulations. Institutional Review Board (IRB) of Washington University (St. Louis) approved all experimental protocols (IRB # 201204036; Title: 'Mapping the Human Connectome: Structure, Function, and Heritability'). Each participant provided written informed consent for analysis of data and publication of results.

**Estimation of cortical thickness and surface area**

MR gradient nonlinearity-induced distortions of the T1-weighted anatomical images were corrected using the gradient_nonlin_unwarp package of Freesurfer(Fischl, 2012) (http://surfer.nmr.mgh.harvard.edu/). Rigid alignment of T1 weighted images, brain extractions and readout distortion corrections were performed using the PreFreeSurfer pipeline of HCP. Output images were internally cropped to remove the neck using the robustfov tool of FSL and those images were non-linearly warped to the MNI space using FMRIB's Nonlinear Image Registration Tool (FNIRT) (Jenkinson, Beckmann, Behrens, Woolrich, & Smith, 2012) (www.fmrib.ox.ac.uk/fsl). After intensity normalization and down-sampling to 1.0 mm isotropic resolution, anatomical images were further processed using the 'recon-all' pipeline of Freesurfer to estimate the thickness and surface areas of all cortical regions defined as per the Desikan-Killiany atlas (Desikan et al., 2006). Estimations of regional cortical thickness and surface area were performed as part of the human connectome project using a minimal processing pipeline (Glasser et al., 2013) and were made available to researchers, which we used for the present analyses.

**Quantitative genetic analyses of cortical morphological phenotypes**
We have used the Sequential Oligogenic Linkage Analysis Routines (SOLAR) (http://solar-

eclipse-genetics.org/) to perform all quantitative genetic analyses of cortical morphological phenotypes (thickness and surface areas). SOLAR uses maximum likelihood variance-decomposition methods which are optimally efficient to furnish maximal genetic information (Almasy & Blangero, 1998). For all genetic analyses described in this sub-section, we have divided the cortical thickness and surface area measures of all cortical regions of each participant by the average cortical thickness and total surface area respectively to control for global effects. Variables were inverse normal transformed to ensure normality and to avoid large residual kurtosis. As per the HCP S1200 data release manual, twins/siblings from the same family do not imply that they had shared household or upbringing environment (HCP, 2017). Further information on whether twins/siblings shared common household/upbringing environment was not available. Therefore, we have not assumed the same household/ upbringing environment for the twins/ siblings from the same family. We have considered age, gender, laterality quotient (handedness measure) as potential covariates. SOLAR automatically includes any covariate correlated with the morphological phenotype of interest at p < 0.10 level during the estimation of morphological phenotypic correlations and in the heritability and genetic correlation estimation models.

Heritability ($h^2 = \sigma_g^2/\sigma_p^2$) represents the proportion of the phenotypic variance ($\sigma_p^2$) attributed to the total additive genetic variance ($\sigma_g^2$). Heritability of the neuroimaging based morphological phenotypes for each cortical region were estimated by comparing the covariance matrix of the morphological phenotype with the covariance matrix predicted by genetic proximity/kinship (Almasy & Blangero, 1998).

Genetic correlations ($\rho_g$) between thickness (or surface area) of all cortical regions were estimated using bivariate polygenic analyses (Almasy et al., 1997) to test whether shared genetic factors influence thickness (or surface area) of different cortical regions. Significant ($p < 0.05$) genetic correlation between thickness (or surface area) measures of two cortical regions would imply that thickness (or surface area) of those two regions were influenced by shared genetic factors. Co-operative genetic influence ($\rho_g > 0$, $p < 0.05$) between the morphological measures of two regions suggests that the shared genetic factors influence change in the morphological phenotypes of both regions in the same directions (e.g., expansion-expansion). On the other hand, competitive genetic influence ($\rho_g < 0$, $p < 0.05$) suggests that the shared genetic factors influence change in the morphological phenotypes in those two regions in opposite directions (expansion-contraction). We constructed inter-regional genetic correlation matrices corresponding to cortical thickness and surface areas (Fig. 1a, 1b). In those matrices, $\rho_g = 0$ corresponds to non-significant ($p \geq 0.05$) (neutral) inter-regional genetic influences. After thresholding of the correlation matrices at the $p < 0.05$ level, the associated cooperative and competitive sub-networks remain connected. However, if p value is further lowered to threshold the genetic correlation matrices, the sub-networks become severely fragmented and isolated nodes appear. We have also constructed correlation matrices for each morphological phenotype (thickness, surface area) using the SOLAR estimates of phenotypic correlation between anatomical regions. We have thresholded those phenotypic correlation matrices at $p < 0.05$ levels.

Further, we estimated the genetic correlation between thickness and surface area of each cortical region using bivariate polygenic analyses (Almasy et al., 1997) to test whether shared genetic factors influence these morphological phenotypes.

**Estimation of signed network properties**

As the genetic correlation matrices have both positive and negative elements, we represented genetic networks underlying each morphological phenotype with a signed graph, $G = (V, E)$ consisting $N$ nodes ($|V| = N$) and edges ($E$). Given that the HCP dataset employed the Desikan-Killiany (Desikan et al., 2006) cortical parcellation scheme, which comprises 34 cortical regions of interest in each cerebral hemisphere, we have 68 nodes in our networks.

$A = A(G)$ is a $N \times N$ signed, weighted, symmetric adjacency matrix without self-loops. The values of off-diagonal elements of *A* could vary between +1 to -1. We estimated the *degree*, *assortativity by degree*, *clustering coefficient*, *modularity* and *participation coefficient* of the signed genetic networks using signed network specific functions of Brain Connectivity Toolbox (Rubinov & Sporns, 2010). We have described these network measures in detail at Supplementary Methods. *Nodal degree*, which reflects the importance of the node in the network, corresponds to the number of links of that node to the other nodes. Mean nodal degree of a network is generally denoted as the density of the network. *Assortativity by degree* estimates the extent to which nodes are connected to nodes of similar degrees. Nodal *clustering coefficients* were estimated considering weighted links and global clustering coefficient were estimated by averaging the corresponding nodal clustering coefficients. The *modularity* measure furnishes information regarding the community structure of a network, with nodes belonging to the same module having strong influence on other nodes within that module and weak influence on the nodes external to that module. The value of nodal *participation coefficient*, a measure of diversity of intermodular links, could vary between zero to one (Guimera & Amaral, 2005). If all the links of a node are restricted within a module which it belongs to, the value of participation coefficient of that node is zero. If the links of the node are uniformly distributed to all modules, the value of participation coefficient of that node would be one. Higher values of the nodal participant coefficient would suggest more diverse association of a node with different modules. We have estimated the degree, assortativity, clustering coefficient and participation coefficients for co-operative and competitive sub-networks. In order to compare the non-trivial properties of a true network to the null model, we generated random networks by simultaneously preserving the degree, weight and strength distributions of the network with positive and negative weights (Rubinov & Sporns, 2011). We have generated the random networks using signed network specific function of the Brain Connectivity Toolbox (Rubinov & Sporns, 2010).

**Estimation of global and local structural balance**

As per the theory of structural balance (Cartwright & Harary, 1956; F. Heider, 1946; Fritz Heider, 1958), in a simple hypothetical network consisting of only three interacting nodes, if all three interactions between the regions are co-operative or if two interactions are competitive and the third interaction is co-operative, the organization of the network would be considered completely global balanced (Fig. 1c). On the other hand, if all three interactions are competitive or if two interactions are cooperative and the third interaction is competitive, then the global network organization is completely imbalanced or conflicted (Fig. 1d). However, in real world networks, the interaction of a node is not limited to just two other nodes and the global organizations of those networks are rarely completely balanced (Estrada & Benzi, 2014). Therefore, instead of conceptualising the structural balance of a network as alternating between two categories of either being completely balanced or being completely imbalanced, the state of global balance of a network can be characterized in terms of the extent to which the network is balanced or imbalanced (Cartwright & Harary, 1956; Estrada & Benzi, 2014).

In the previous sub-section, we have mentioned the signed adjacency matrix $A$, underlying the graph $G$. Here, we introduce the unsigned adjacency matrix $|A| = A(|G|)$ of the unsigned graph $|G|$. As interactions of a node in real-world network may not be limited to only two other nodes, we followed a walk-based approach to characterize the state of balance of the signed networks where more importance was given to the shorter cycles than the longer ones (Estrada & Benzi, 2014). The sign of a closed walk (sequences of vertices starting and ending at same vertex) on a signed graph ($G$) is the sign of the product of the signs of the links associated with the walk. A balanced (or imbalanced) weighted closed walk of length greater than zero has a positive (or negative) sign. The extent of lack of balance in a signed network can be defined by the ratio of sum of weighted imbalanced and balanced closed walks (Estrada & Benzi, 2014),

$$U = \frac{1-K}{1+K}$$

where $U$ in the measure of global imbalance of the network. If the network is completely imbalanced, $U = 1$ and completely balanced when $U = 0$. $K$ is defined as,

$$K = \frac{\sum_{i=1}^{N} \exp(\lambda_i)}{\sum_{i=1}^{N} \exp(\mu_i)}$$

where $\lambda_i$ and $\mu_i$ are the $i$-th eigenvalues of $A$ and $|A|$ respectively. Similarly, nodal/local imbalance measure for the $i$-th node is defined as (Estrada & Benzi, 2014),

$$U_i = \frac{1-K_i}{1+K_i}$$

where, $K_i = [\exp(A)]_{ii} / [\exp(|A|)]_{ii}$. The node is completely imbalanced if $U_i = 1$ and completely balanced when $U_i = 0$. We have estimated the global ($U$) and nodal imbalance ($U_i$) measures to test the extent to which the state of collective regulation of cortical morphological phenotypes by co-operative and competitive genetic influences are globally and locally imbalanced/conflicted.

**Principal component analyses**

We have performed principal component analyses of morphological phenotypic correlation matrices to estimate the principal components and the proportion of variance of morphological phenotypes within the participants that was explained by each of those principal components. If a principal component explained more than 1% variance of thickness (or surface area) within the participants, we have performed Spearman's rank correlation between the loadings of that principal component and the genetic network-based estimates of nodal structural imbalance underlying cortical thickness (or surface area) that we computed earlier.

**Statistical analyses**

We have used a permutation-based statistical framework to compare the network properties of a true network to the null model. We generated 20,000 random networks corresponding to a true network and estimated the p-value by computing the number of network properties derived from the random networks which were greater/lesser than the value of the corresponding true network properties, divided by the total number of random networks. Using the PALM (Permutation Analysis of Linear Models) tool of FSL, we compared the mean nodal measures between the groups involving 10,000 sign-flippings and without assuming equal group variances (Winkler, Ridgway, Webster, Smith, & Nichols, 2014). To test the association between the two nodal network measures, we have ranked the data related to each network measure and estimated the Spearman's rank correlation coefficient ($\rho$). Significances of these statistical tests were determined at $p < 0.05$ level.

## ACKNOWLEDGEMENT

Data were provided by the Human Connectome Project, WU-Minn Consortium (Principal Investigators: David Van Essen and Kamil Ugurbil; 1U54MH091657) funded by the 16 NIH Institutes and Centers that support the NIH Blueprint for Neuroscience Research; and by the McDonnell Center for Systems Neuroscience at Washington University.

## FINANCIAL SUPPORT

SB has been funded by the National Institute for Health Research (NIHR), UK through a Clinician Scientist award (NIHR-CS-11-001). SP is supported by the Newton International Fellowship from the Newton Fund. The views expressed are those of the authors and not necessarily those of the NHS, the NIHR or the Department of Health. The funders had no role in the design and conduct of the study; collection, management, analysis, and interpretation of the data; preparation, review, or approval of the manuscript; and decision to submit the manuscript for publication. All authors have approved the final version of the paper.

## CONFLICT OF INTEREST

None.
20

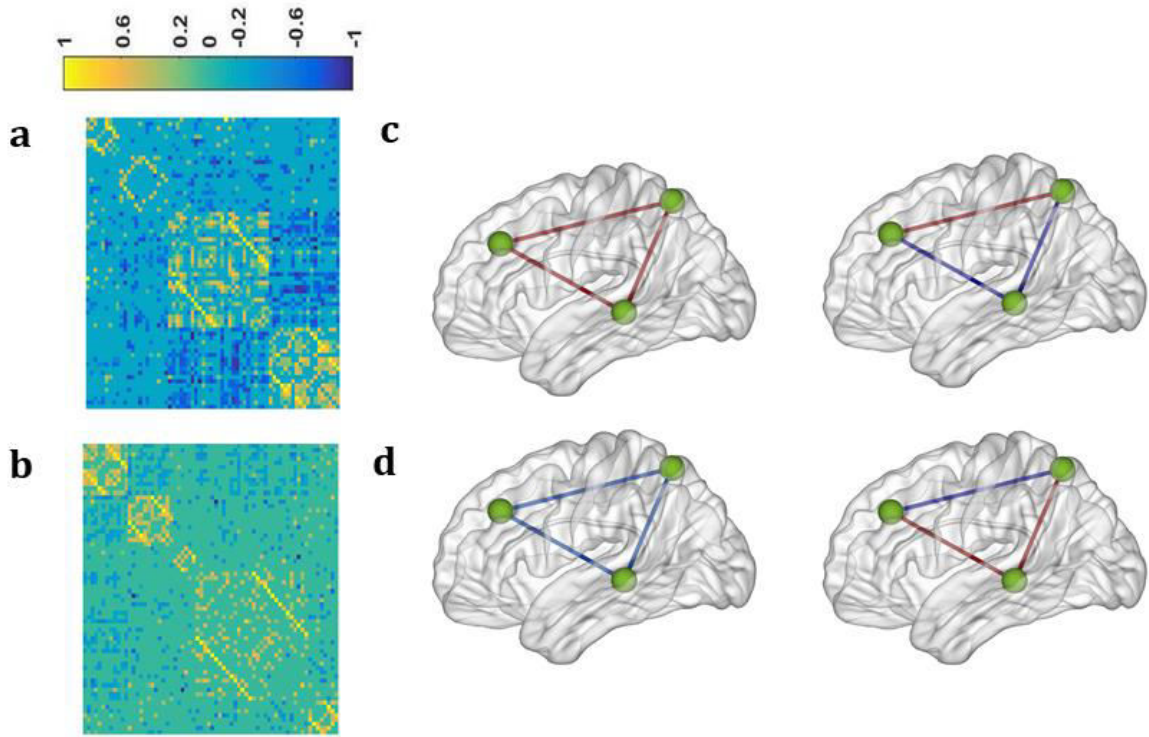

**Figure 1. Genetic correlations and hypothetical networks.** Inter-regional genetic correlations underlying cortical thickness and surface area are shown in panels a and b respectively. Cooperative and competitive influences between three green spherical nodes of a hypothetical network are shown using red and blue lines respectively (c, d). The balanced and unbalanced organization of the network are depicted in panels c and d respectively.

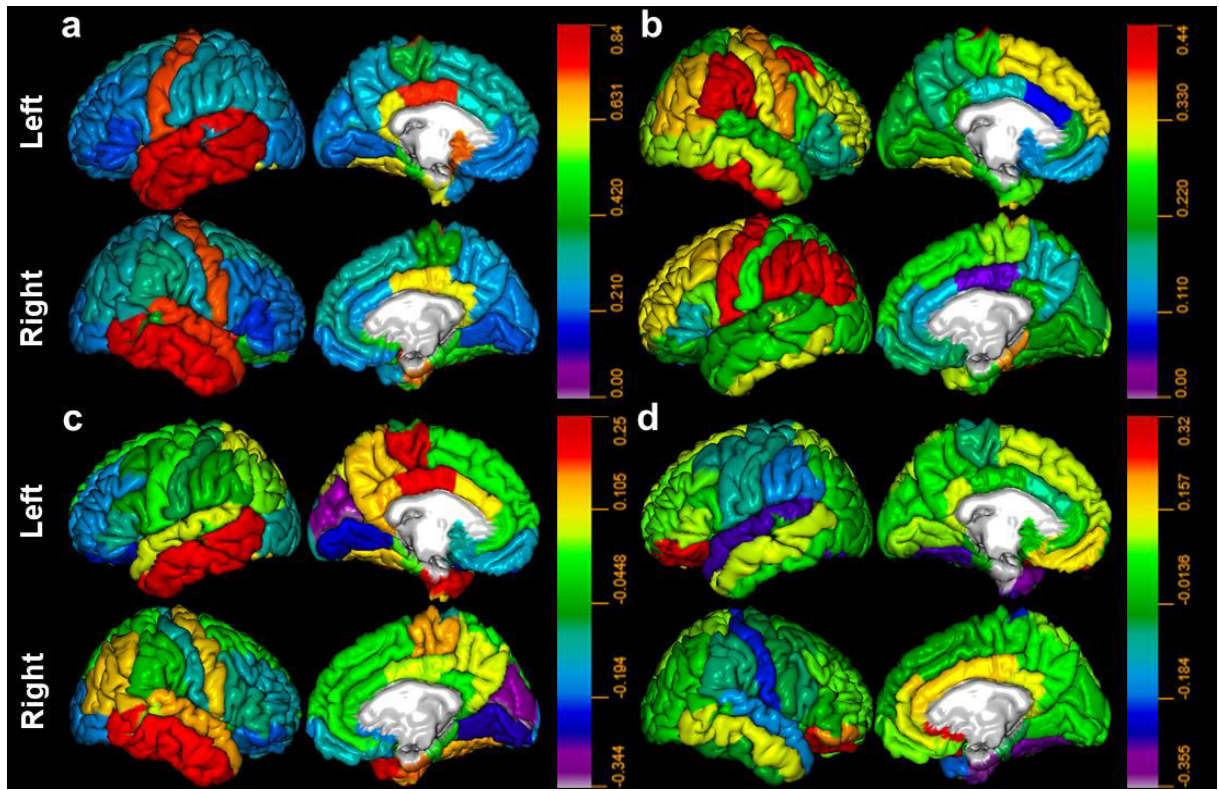

**Figure 2. Structural imbalance maps of cortical genetic networks and principal components of phenotypic variations**. Nodal values of structural imbalance of genetic networks underlying cortical thickness and surface area are displayed in panels a and b respectively. Higher values imply higher structural imbalance. Loadings of principal component 5 underlying cortical thickness and surface area variations within the participants are displayed in panels c and d respectively. Higher absolute values of loadings imply higher contribution of the region in that principal component. Left and right marked in the panels correspond to the left and right hemispheres respectively.

**Table 1.**

**Global network properties of co-operative and competitive cortical genetic influences on thickness and surface area.**

| Global measures | Genetic influences (cortical thickness) | | | Genetic influences (surface area) | | |
|---|---|---|---|---|---|---|
| | True | Random | p | True | Random | p |
| Assortativity (cooperative influences) | 0.525 | -0.087 | $5 \times 10^{-5}$ | 0.177 | -0.0440 | $5 \times 10^{-5}$ |
| Assortativity (competitive influences) | -0.071 | -0.082 | $5 \times 10^{-5}$ | -0.139 | -0.099 | $5 \times 10^{-5}$ |
| Clustering coefficient ($C$) (cooperative a influences) | 0.296 | 0.127 | $5 \times 10^{-5}$ | 0.221 | 0.052 | $5 \times 10^{-5}$ |
| Clustering coefficient ($C$) (competitive influences) | 0.014 | 0.086 | $5 \times 10^{-5}$ | 0.020 | 0.071 | $5 \times 10^{-5}$ |
| $C_{true}/C_{random}$ (cooperative influences) | 2.331 | - | - | 4.250 | - | - |
| $C_{true}/C_{random}$ (competitive influences) | 0.163 | - | - | 0.282 | - | - |
| Modularity ($Q$) | 0.749 | 0.348 | $5 \times 10^{-5}$ | 0.783 | 0.438 | $5 \times 10^{-5}$ |
| $Q_{true}/Q_{random}$ | 2.152 | - | - | 1.789 | - | - |
| Imbalance measure ($U$) | 0.467 | 0.992 | $5 \times 10^{-5}$ | 0.429 | 0.873 | $5 \times 10^{-5}$ |
| $U_{true}/U_{random}$ | 0.471 | - | - | 0.491 | - | - |

$C_{true}$, $Q_{true}$ and $U_{true}$ are the global clustering coefficient, modularity and global imbalance measures of the true networks respectively. $C_{random}$, $Q_{random}$ and $U_{random}$ correspond to the mean global clustering coefficient, mean modularity and mean global imbalance measures of the random networks.

**Table 2**

**Nodal measures of co-operative and competitive genetic influences**

| Nodal measures | Genetic influences (cortical thickness) | | | Genetic influences (surface area) | | |
|---|---|---|---|---|---|---|
| | co-operative (mean ± std) | competitive (mean ± std) | p | co-operative (mean ± std) | competitive (mean ± std) | p |
| Degrees | 10.50±5.60 | 10.90±4.93 | 0.61 | 6.97±2.57 | 8.24±4.80 | 0.36 |
| Clustering coefficient | 0.29±0.13 | 0.01±0.03 | 0.0001 | 0.22±0.10 | 0.02±0.02 | 0.0001 |
| Participation coefficient | 0.22±0.20 | 0.34±0.24 | 0.0013 | 0.21±0.18 | 0.45±0.19 | 0.0001 |

**Table 3**

**Correlations between nodal properties of genetic networks underlying cortical thickness and surface area**

| Nodal properties | Co-operative | | Competitive | |
|---|---|---|---|---|
| | ρ | p | ρ | p |
| Degree | 0.13 | 0.30 | 0.24 | 0.06 |
| Assortativity | -0.089 | 0.47 | 0.13 | 0.30 |
| Clustering coefficient | 0.16 | 0.19 | 0.01 | 0.93 |
| Participation coefficient | -0.09 | 0.45 | 0.10 | 0.40 |

ρ corresponds to Spearman's rank correlation coefficient

# Supplementary Information

## Supplementary Methods

**Assortativity by degree:** Nodal assortativity of the *i*-th node of an undirected network is defined as (Piraveenan, Prokopenko, & Zomaya, 2010):

$$r_i = \frac{1}{2M\sigma_q^2} k(k+1)(\bar{k} - \mu_q)$$

If we reach a vertex following a randomly chosen edge in a network with total *M* edges, the remaining degree (*k*) of the vertex corresponds to the number of links of the vertex except the link we followed to arrive at that vertex. $e_{ij}$ is the joint probability distribution of the remaining degrees of two vertices at either end of a randomly chosen link. $q_j$ is the normalized distribution of remaining degrees, $\mu_q$ and $\sigma_q$ are the mean and standard deviation of this distribution. $\bar{k}$ is the average remaining degree of the node's neighbours. Global assortativity values of the co-operative and competitive genetic sub-networks are the sum of nodal assortativity values of the corresponding sub-networks.

**Clustering coefficient:** Nodal clustering coefficient of the *i*-th node of a weighted undirected network can be defined as (Zhang & Horvath, 2005):

$$C_i = \frac{n_i}{\pi_i}$$

$C_i$ of a node ranges between 0 and 1. where

$$n_i = \frac{1}{2} \Sigma_{u \neq i} \Sigma_{\{v|v \neq i, v \neq u\}} a_{iu} a_{uv} a_{vj}$$

$a_{ij}$ is link weight ($0 < a_{ij} < 1$) and

$$\pi_i = \frac{1}{2}((\Sigma_{u \neq i} a_{iu})^2 - \Sigma_{u \neq i} a_{iu}^2)$$

We have estimated the nodal clustering coefficients ($C_i$) for co-operative and competitive genetic sub-networks. For competitive sub-networks, we have considered the absolute values of the link weights. Global clustering coefficients of those sub-networks were estimated by averaging the corresponding nodal clustering coefficients.

**Modularity and Participation Coefficient:**

For community detection and estimation of modularity *values* of signed networks, we have used Brain Connectivity Toolbox based multi-resolution Louvain community detection algorithm with additional iterative modularity maximization and by treating the positive and negative weights asymmetrically (Rubinov & Sporns, 2011). After determining the modules, we have estimated the nodal participation coefficients for co-operative and competitive sub-networks. Nodal participation coefficient of node *i* is defined as (Guimera & Amaral, 2005):

$$PC_i = 1 - \Sigma_{n=1}^{N_M}\left(\frac{k_{is}}{k_i}\right)^2$$

where $k_{is}$ is the number of links of the node $i$ to the nodes in module $s$ and $k_i$ is the total number of links of node $i$. $N_M$ is the total number of modules in the signed network.

# Supplementary References

**Supplementary Figure**

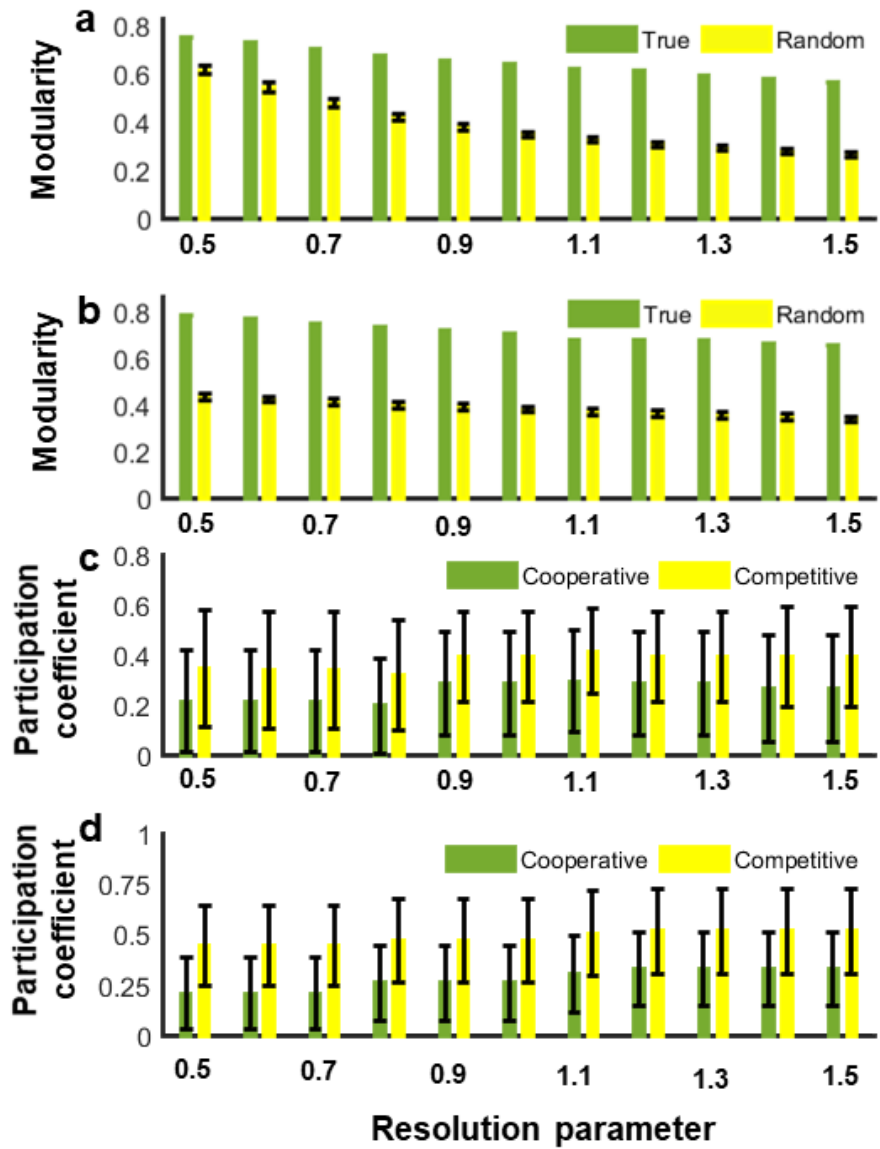

**Supplementary Figure 1. Modularity and Participation Coefficient at different values of resolution parameter.** Modularity values of genetic networks underlying thickness and surface area are displayed in panels a and b respectively. Heights of the yellow bars and error bars represent means and standard deviations of modularity of random networks (a, b). True networks are more modular than random networks (p = 0.00001). Means and standard deviations of cooperative and competitive genetic influences underlying thickness and surface area are displayed in panels c and d respectively. Heights of the bars and error bars represent means and standard deviations. Mean values of participation coefficient of competitive genetic influences are significantly higher than cooperative influences (p < 0.004). Resolution parameter was varied from 0.5 to 1.5.